\documentclass{aa}
\usepackage{times}
\usepackage{graphics}
\usepackage{epsfig}
\begin{document}
\thesaurus 
{08(08.23.1,08.13.1)}
\title{{\em Letter to the Editor} \\
  {Four magnetic DB white dwarfs discovered by the}\\ 
  {Hamburg/ESO survey}\thanks{Based on observations
   at the European Southern Observatory, La Silla, Chile}}
\author {D. Reimers\inst{1}  
\and S. Jordan\inst{2}
\and V. Beckmann\inst{1}
\and N. Christlieb\inst{1}
\and L. Wisotzki\inst{1}
}
\institute{Hamburger Sternwarte, Universit\"at Hamburg, Gojenbergsweg
  112, D-21029 Hamburg, Germany
\and Institut f\"ur Theoretische Physik und Astrophysik der Universit\"at
  Kiel, D-24098 Kiel, Germany}
\offprints{D.~Reimers, dreimers@hs.uni-hamburg.de}
\date{received 30.6.98; accepted 9.7.98}
\titlerunning{Discovery of four magnetic DB white dwarfs}
\authorrunning{D. Reimers et al.}
\maketitle
\begin{abstract}
  We report on seven peculiar faint blue stars found in the course of the
  Hamburg/ESO survey (HES) which appear to be magnetic white dwarfs (WDs) with
  non-hydrogen spectra. We show in particular that four of them
  (\object{HE~0338--3853}, \object{HE~0107--0158}, \object{HE~0026--2150}, and
  \object{HE~0003--5701}) have \ion{He}{I} lines split by magnetic fields of
  roughly 20\,MG, since the $\pi$ components of \ion{He}{I} 5876\,{\AA} and
  \ion{He}{I} 4929\,{\AA} can be identified unambiguously in their spectra,
  and the $\sigma^+$, $\sigma^-$ components can be identified in the spectra
  of two of these stars (\object{HE~0338--3853} and \object{HE~0003--5701}).
  Besides \object{GD~229}, these are the first magnetic DB white dwarfs
  discovered so far.

  In addition, three further WDs with broad, unidentifiable features have been
  found: \object{HE~1043--0502}, \object{HE~0236--2656}, and
  \object{HE~0330--0002}. We argue that in all three of these stars \ion{H}{I}
  can not be responsible for the broad features, and \ion{He}{I} most
  probably not for the features in \object{HE~0236--2656} and
  \object{HE~0330--0002}, while it still remains possible that the
  broad features of \object{HE~1043--0502} are due to \ion{He}{I}.
  
\end{abstract}

\section{Introduction}
Compact stars with strong magnetic fields are rewarding objects
to study. They offer the opportunity to probe atoms under the
influence of strong magnetic fields not available in laboratory
experiments. Furthermore, magnetic WDs should give information on
the origin and evolution of stellar magnetic fields from the
main sequence (e.g. Ap stars) to the final evolutionary stages.

For reviews of the subject, cf. Angel (1978), Koester \& Chanmugam (1990),
Chanmugam (1992), Schmidt (1995), and Jordan (1997). Jordan lists 50 magnetic
WDs, but only 15 with extemely high field strengths ($>100$\,MG).  Nearly all
magnetic WDs have pure hydrogen spectra, while, surprisingly, magnetic DBs
have been missing so far. Until very recently, the only known magnetic WD with
unambiguously identified \ion{He}{I} lines was Feige~7, which shows strong,
magnetically split hydrogen {\em and\/} \ion{He}{I} lines (Achilleos et al.
1992). However, among the high magnetic field WDs there were two stars with
strong unidentified features, namely \object{GD~229} and \object{LB~11146B}.
The latter has a broad absorption feature at 5500\,{\AA}, and in addition a
Ly$\alpha\,\sigma$ component at 1340\,{\AA} and a stationary H$\alpha$
component at $\sim\,8500$\,{\AA}, suggesting a high field strength of 670\,MG
(Liebert et al. 1993; Glenn at al. 1994). It was suggested that since the
strong features in \object{GD~229} and \object{LB~11146B} cannot be hydrogen,
the next likely atmospheric constituent, namely He, is responsible for the
broad features. Very recently Jordan et al. (1998) indeed have succeeded in
identifying most of the \object{GD~229} features as due to \ion{He}{I} in
fields between 300 and 700\,MG.

The Hamburg/ESO survey for bright QSOs has turned out to be a rich
source of magnetic WDs (Reimers et al. 1994, 1996). Due to the high spectral
resolution ($\sim\,10\dots 20$\,{\AA} FWHM) of the HES objective prism plates
taken with the ESO Schmidt-telescope, the stellar component with identifiable
features (H, He, \ion{Ca}{II},\dots) can be removed from the QSO candidate
sample (Wisotzki et al. 1996; cf.  Reimers \& Wisotzki, 1997, for further
references). The fraction of non-QSOs within the sample of quasar candidates
for which follow-up spectroscopy is finally performed therefore contains
considerable numbers of peculiar blue stars; in particular magnetic WDs, CVs,
and binaries. In addition to the stars already published we have roughly 20
further magnetic WD candidates, most of them with hydrogen spectra.

In this letter we announce the discovery of seven new peculiar white
dwarfs; four of them are magnetic DBs with detectable Zeeman lines.

\section{Observations}

\begin{figure*}
  \begin{center}
    \epsfig{file= 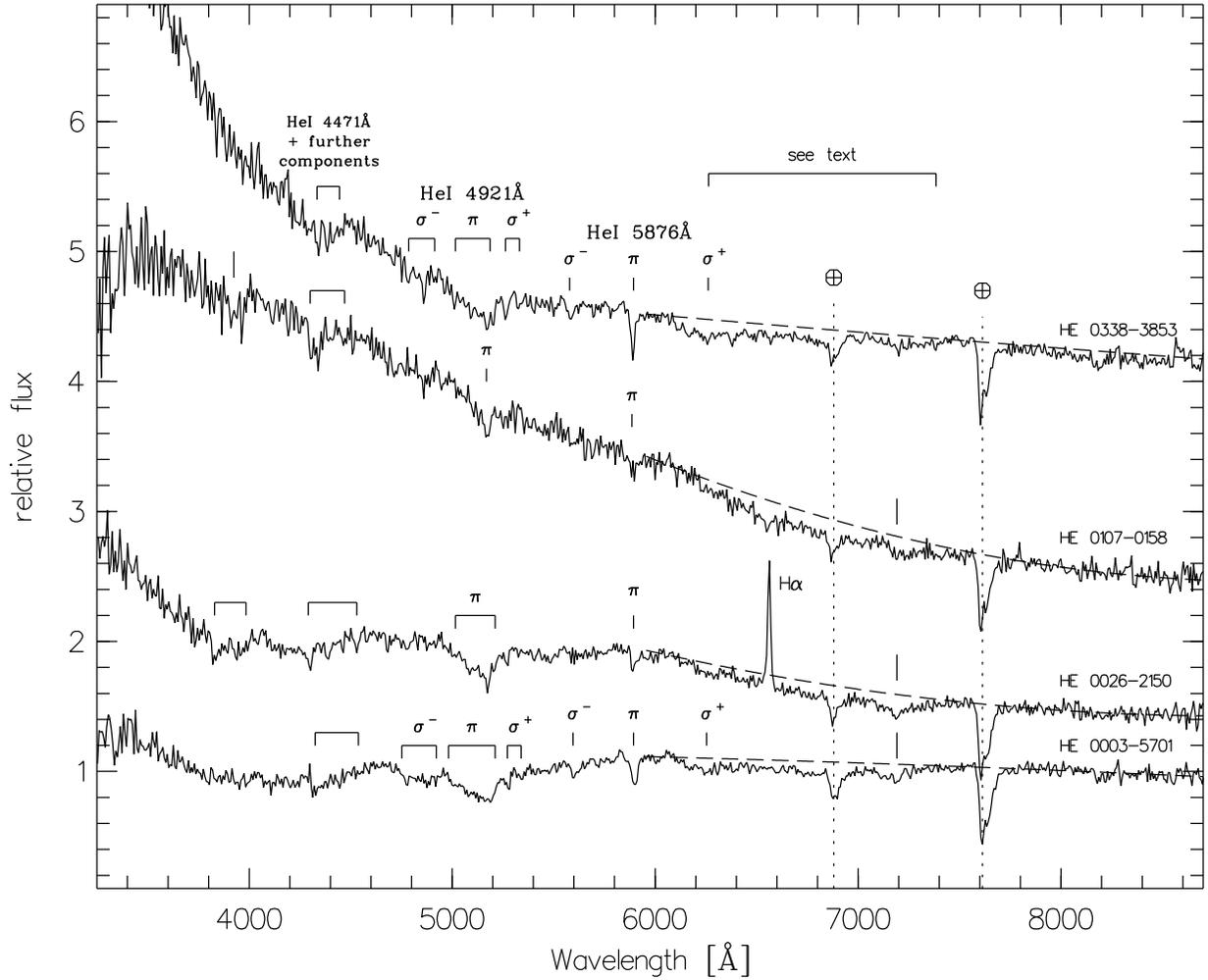, clip=, width=16.5cm,
      bbllx=72, bblly=392, bburx=540, bbury=772}
  \end{center}
  \caption{\label{magDBs} Flux spectra $f_\lambda$ of
    the four magnetic DBs taken with the ESO 1.52\,m telescope, together with
    a rough continuum estimation in the red part of the spectra. The continua
    have been determined by fitting a second order polynomial through some
    high points, and are intended to make the shallow, broad absorption
    features in the red region noticeable. Atmospheric bands are labeled with
    `$\oplus$'. All spectra have been normalized to a mean flux of unity in
    the region 7000--7500\,{\AA}, and afterwards arbitrarily shifted in
    $y$-direction.  Though these snapshot-spectra are nonphotometric, their
    apparent temperature sequence is consistent with colours estimated from the
    digitized objective prism spectra.}
\end{figure*}

\begin{figure*}
  \begin{center}
    \epsfig{file= 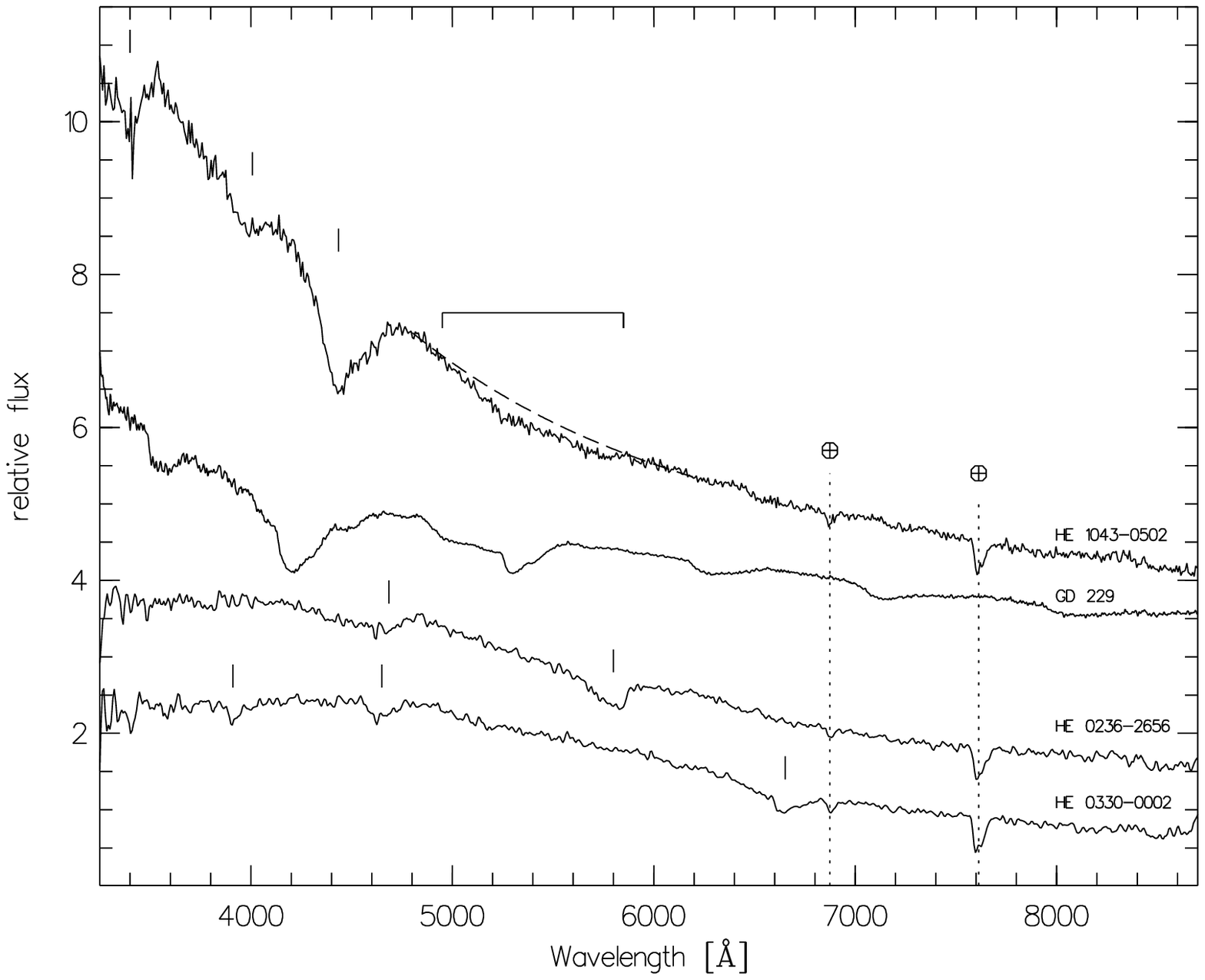, clip=, width=16.5cm,
      bbllx=68, bblly=392, bburx=536, bbury=772}
  \end{center}
  \caption{\label{unidWDs} Flux spectra of three WDs with
    unidentifiable features in comparison with a spectrum of \object{GD~229}.
    The latter is the same as in Schmidt et al. (1996) and was kindly provided
    by G. Schmidt. The spectra have been treated in the same way as descripted
    in the caption of Fig. \ref{magDBs}. The spectra of \object{HE~0236--2656}
    and \object{HE~0330--0002} have been smoothed with a gaussian filter of
    15\,{\AA} FWHM. A rough continuum estimation of the 5000--5800\,{\AA}
    region by a third order polynomial is given for \object{HE~1043--0502} to
    visualize the shallow, broad absorption feature. Telluric bands are
    labeled with `$\oplus$'. As for the spectra shown in Fig. \ref{magDBs}, the
    apparent temperature sequence is consistent with colours estimated from
    the digitized objective prism spectra.}
\end{figure*}

Coordinates and magnitudes of the seven objects are given in Table
\ref{coordsmag}.  The coordinates are accurate to $\pm 0.5\arcsec$. No finding
charts are given since the objects can be localized unambiguously using the
{\em Digital Sky Survey}. Details of the spectroscopic follow-up observations
are presented in Table \ref{observations}.  Spectra are shown in Fig.
\ref{magDBs} and Fig. \ref{unidWDs}, together with a spectrum of
\object{GD~229} from Schmidt et al.  (1996).

\begin{table}[htbp]
  \caption[]{\label{coordsmag} Coordindates and magnitudes of the
    objects. The $B_J$ band is defined by the sensitivity curve of the
    hyper-sensitized Kodak IIIa--J emulsion, folded with the filter function
    of a Schott BG395 filter. $B_J$ is roughly equal to $B$ for objects around
    $B-V=0$. The listed $B_J$ magnitudes are accurate to $\pm 0.1$\,mag.
    }
  \begin{flushleft}
    \begin{tabular}{lllll}\hline
      Object & Type & R.A. (2000.0) & Decl. & $B_J$ \rule{0.0ex}{2.5ex}\\\hline
      \object{HE~0338--3853} & mag. DB & 03:40:36.8 & $-$38:43:43 &
      16.7\rule{0.0ex}{2.5ex}\\  
      \object{HE~0107--0158} & mag. DB & 01:10:13.2 & $-$01:42:03 & 16.4 \\
      \object{HE~0026--2150} & mag. DB & 00:29:26.1 & $-$21:33:42 & 16.6 \\
      \object{HE~0003--5701} & mag. DB & 00:05:39.2 & $-$56:44:41 & 15.8 \\
      \object{HE~1043--0502} & unid. & 10:46:09.8 & $-$05:18:17 & 17.0 \\
      \object{HE~0236--2656} & unid. & 02:38:41.2 & $-$26:43:24 & 16.7 \\
      \object{HE~0330--0002} & unid. & 03:33:20.2 & $+$00:07:20 & 16.8 \\\hline
    \end{tabular}
  \end{flushleft}
\end{table}

\begin{table}[htbp]
  \caption[]{\label{observations} Journal of observations. U.T. is given
    for the start of the exposure. All objects have been observed with the
    B\&C spectrograph attached to the ESO 1.52\,m telescope, at a spectral
    resolution of $\sim 15$\,{\AA}.}
  \begin{flushleft}
    \begin{tabular}{lllr}\hline
      Object & Date & U.T. & \multicolumn{1}{l}{Exp. time}
      \rule{0.0ex}{2.5ex}\\\hline
      \object{HE~0338--3853} & 02/11/95 & 6:31 & 20\,min\rule{0.0ex}{2.5ex}\\  
      \object{HE~0107--0158} & 23/10/97 & 5:52 &  8\,min\\
      \object{HE~0026--2150} & 23/10/97 & 1:52 & 10\,min\\
      \object{HE~0003--5701} & 05/10/96 & 1:40 &  5\,min\\
      \object{HE~1043--0502} & 05/02/97 & 5:20 & 20\,min\\
      \object{HE~0236--2656} & 06/12/97 & 4:12 &  5\,min\\
      \object{HE~0330--0002} & 04/12/97 & 4:57 &  5\,min\\\hline
    \end{tabular}
  \end{flushleft}
\end{table}

It should be emphasized that all spectra have been taken in the course of
quasar candidate follow-up spectroscopy and that typically the
spectrophotometry of these snapshot spectra is not sufficiently accurate for
determining temperatures of the stars. We notice, however, that the apparent
temperature sequence with \object{HE~0338--3853} and \object{HE~1043--0502} as
being the hottest stars while \object{HE~0236--2656} and
\object{HE~0330--0002} are the coolest among the seven stars is confirmed by
the colours estimated from the digitized objective prism spectra.

\section{Interpretation}
\subsection{Magnetic DB white dwarfs}

Four of the new WDs have remarkably similar spectra (Fig. \ref{magDBs}). In
two of them (\object{HE~0338--3853} and \object{HE~0003--5701}) the features
can be identified with \ion{He}{I} lines split under magnetic fields of
roughly 20\,MG. As in case of Feige~7 (Achilleos et al. 1992) the central
$\pi$ component of \ion{He}{I} 5876\,{\AA} as well as the $\sigma^+$ and
$\sigma^-$ components are clearly detected. The broad feature between 5000 and
5200\,{\AA} consists of the five $\pi$ components of \ion{He}{I} 4929\,{\AA}.
The corresponding blue shifted $\sigma^-$ components are unresolved and are
visible as broad absorption at $\sim 4800$\,{\AA}, while the two $\sigma^+$
components are seen at roughly 5270\,{\AA} and 5340\,{\AA}, respectively.
Comparison with the tables of Kemic (1974), and also Fig. 3 in Achilleos et
al. (1992), hints at field strengths in the range of 20 to 25\,MG. The
additional features between roughly 6200\,{\AA} and 7300\,{\AA} are probably
due to the Zeeman-triplets of \ion{He}{I} 6678\,{\AA} and \ion{He}{I}
7065\,{\AA}. According to the new computations of Schmelcher (priv. comm.),
the two triplets cover the range longward of 6200\,{\AA} at field strengths of
20\,MG. These features are visible as a broad, shallow absorption trough with
evidence for individual lines, e.g. at 7200\,{\AA} (Fig. \ref{magDBs}). At
6200\,{\AA} there is a blending with the $\sigma^+$ component of \ion{He}{I}
5876\,{\AA}.  At higher spectral resolution and $S/N$ the individual triplett
components (6 lines) can probably identified. The features in the blue part of
the spectra can also be attributed to He lines at similar field strengths;
partly to \ion{He}{I} 4471\,{\AA}. So far, there is no evidence for hydrogen
lines.

The other two stars have quite similar spectra, although the features are
weaker (\object{HE~0026--2150}), or the spectrum is too noisy to show narrow,
weak features (\object{HE~0107--0158}). However, the gross appearance of the
stronger features is rather similar: The shifted $\pi$ components of
\ion{He}{I} 5876\,{\AA} and in particular of \ion{He}{I} 4929\,{\AA} at
5100\,{\AA} are detected. Both stars apparently have \ion{He}{I} spectra split
by magnetic fields in the range of 10 to 30\,MG. \object{HE~0026--2150} has
an additional strong, sharp H$\alpha$ emission line and is probably a binary.


\subsection{Stars with unidentifiable features}

Besides the magnetic DBs we have found three further stars with peculiar
spectra (Fig. \ref{unidWDs}), probably also magnetic non-DAs.
\object{HE~1043--0502} has a strong feature at 4430\,{\AA} similar to the
4300\,{\AA} trough in \object{GD~229}. Further features can be seen roughly at
3400\,{\AA} and 4000\,{\AA}. In addition, there is an extremely broad, shallow
trough between 5000\,{\AA} and 5800\,{\AA}.

Although the spectra of \object{HE~1043--0502} and \object{GD~229} look
similar at first glance, Jordan et al. (1998) on the one hand have {\em
  succeeded\/} in identifying most of the features of \object{GD~229} as
stationary \ion{He}{I} transitions in a field between 300 and 700\,MG, while
on the other hand there is {\em no\/} possible match of the features in the
spectrum of \object{HE~1043--0502} with \ion{He}{I} in magnetic fields of up
to $2\cdot 10^3$\,MG (Becken \& Schmelcher, priv comm.; cf. Jordan et al.
1998).  However, even in the case of GD\,229 the strongest features at
$4000-4200$\,\AA\ and at about $5280$\,\AA\ can only be attributed to two line
transitions ($2^1 0^+\rightarrow 2^1 0^-$, $2^1 0^+\rightarrow 2^1 (-1)^+$),
while other and weaker ones are produced by a superposition of several
stationary components.  Therefore, Jordan et al. (1998) speculated that
additional stationary line components must be present which were not included
in the current data sets for \ion{He}{I} in a strong magnetic field. All
available data are restricted to transitions with magnetic quantum numbers
$|m|\le 1$.


We have also no identification of the shallow features in the spectra of
\object{HE~0236--2656} (at 4650\,{\AA} and 5800\,{\AA}) and
\object{HE~0330--0002} (at 3900\,{\AA}, 4650\,{\AA}, and 6650\,{\AA}). In the
former, the broad absorption troughs could be molecular bands according to
their form, but we do not have an identification. Apart from missing
identifications, the apparently low temperature of these two objects argues
against the presence of \ion{He}{I} lines. We tentatively conclude that
neither hydrogen nor helium give a plausible identification of the features.

\section{Final remarks}

Given the fact that except \object{GD~229} there is no DB in the sample of 50
magnetic WDs known today (Jordan 1997), our discovery of four magnetic DBs
appears surprising.  The question is: Why have no magnetic DBs been discovered
in the course of proper motion surveys or the PG survey, with exception of
\object{GD~229}? The probable reason is the rareness of the phenomenon.  With
respect to the detection of peculiar blue stars, the HES should be complete to
$B_J\sim 17$.  The HES follow-up observations of candidates from 280 ESO/SRC
fields, corresponding to an effective area of $\sim 5000$ square degrees, have
revealed 4 magnetic DBs. Only \object{HE~0003--5701} is significantly brighter
than the PG magnitude limit ($B_J=16.5$). On the same area roughly three
dozens DBs have been found by the HES (unpublished). Although we are aware of
the fact that this is still small number statistics, we tentatively conclude
that the incidence of magnetism is not lower in DBs than in DAs.

Finally, we wish to point out that most of our results are preliminary. The
main intention of this letter is to draw the attention of the white dwarf
research community to these objects.

{\em Acknowledgements.} This work has been supported by grants from the
Deutsche Forschungsgemeinschaft under Re 353/39--1, Re 353/40--1 and Ko
738/7--1.  We are particulary grateful to P. Schmelcher, Heidelberg, for the
communication of his unpublished \ion{He}{I} calculations.

\end{document}